\documentclass[12pt,thmsa]{article}
\usepackage{sw20lart}

\input{tcilatex}
\begin{document}

\title{Quantum mechanics vs. local realism, is that the question?}
\author{Emilio Santos \\
%EndAName
Departamento de F\'{i}sica, Universidad de Cantabria, Santander, Spain}
\maketitle

\begin{abstract}
The conjecture is made that quantum mechanics is compatible with local
hidden variables (or local realism). The conjecture seems to be ruled out by
the \textit{theoretical }argument of Bell, but it is supported by the 
\textit{empirical }fact that nobody has been able to perform a loophole-free
test of local realism in spite of renewed effort during almost 40 years.
\end{abstract}

During the last three decades many experiments have been performed aimed at
ruling out any local hidden variables (LHV) theory. In spite of the effort
some deficiencies in the proof remain, the most important being described by 
\textit{locality }and\textit{\ detector efficiency }loopholes. According to
recent reports the two loopholes have been closed \cite{Weihs}\cite{Rowe}.
However, each loophole has been closed in a different experiment. That this
is not enough to refute all LHV theories has been pointed out by Vaidman 
\cite{Vaidman}. I agree with that paper except for a point which will be
commented below. Therefore I shall not spent time in repeating the
arguments, which may be summarized as follows: The existence of a loophole
means that quantum mechanics and LHV are actually compatible for the
experiment and, therefore, the experiment is unable to discriminate between
quantum mechanics and LHV theories. The fact that quantum predictions are
verified, certainly reinforces our belief in the correctness of quantum
theory. But the experiment says nothing against LHV theories.

The opinion of Vaidman, and many other people, with which I disagree is
expressed in his sentence: ''there is no real question what will be the
outcome of this type of (loophole-free) experiments: the predictions of
quantum theory or results conforming with the Bell inequalities. ...only a
minute minority of physicists believe that quantum mechanics might fail in
this type of experiments.'' I would not disagree with the sentence if I were
willing to accept the alternative ''either quantum mechanics or LHV theories
'', but I do not accept it, as explained below.

After the discovery of the steam engine in the XVIII Century, many people
attempted to construct a ''perpetuum mobile'', that is a machine able to
work for ever either without energy supply or extracting energy by just
cooling the environment. They failed, and this lead physicists to postulate
that the ''perpetuum mobile'' is impossible, which on turn is the basis for
the principles of thermodynamics. Now, for almost 40 years, many people have
tried to perform a loophole-free Bell test, and they have failed. (By way of
comparison we may recall that after the proposal by Lee and Yang that parity
is not always conserved, this was proved in an uncontroversial
(loophole-free) experiment by Wu et al. in less than one year). Consequently
it is not so absurd if I conjecture that loophole-free Bell tests are
impossible.

If we leave outside theoretical arguments, we have two empirical facts: 1)
The predictions of quantum mechanics have been verified in many experiments
with unprecedent precision, 2) No loophole-free Bell test has yet been
performed. Therefore there is \textit{a real open question}, namely to
confirm or disprove the conjecture above stated. A single loophole-free
experiment violating a Bell inequality would show that the conjecture is
false. But after every attempt at performing such an experiment fails the
conjecture becomes reinforced. This is similar to the reinforcement of
quantum theory after every experiment that verifies its predictions.

The compatibility between quantum mechanics and LHV theories, not yet
disproved \textit{empirically, }contradicts the \textit{theoretical }%
argument known as Bell\'{}s theorem. Therefore, might be the case that
Bell\'{}s theorem is false? In order to answer this question we firstly
remember that any theorem is a \textit{mathematical} statement whose
relation with empirical facts is not straightforward in general. In quantum
mechanics there are two quite different ingredients, namely the formalism
(that is the equations) and the (''semantical'') rules for the conection
with experiments (for instance, the assumption that all \textit{states}
which may be actually prepared in the laboratory can be represented by 
\textit{density operators} on Hilbert space). I \textit{do believe} that the
equations of quantum theory are correct, but I think that the standard
semantical rules may be questioned. It is the case that the most spectacular
verifications of quantum theory (e.g. the value of the Lamb shift or the
magnetic moment of the electron) depend strongly on the equations but very
weakly on the semantical rules. In sharp contrast, the proof of Bell's
theorem depends strongly on the semantical rules. I explain the point in
more detail in the following.

The standard proof of Bell's theorem requires assuming the existence of the
singlet state of two spin-1/2 particles. Quantum theory predicts a violation
of the Bell inequality if the spin projections of the particles in this
state are measured at spacelike separation. But, without changing in any way
the quantum formalism, we might assume that the singlet pure state cannot be
prepared in the laboratory. For instance we may suppose that only those
states of two spin-1/2 particles (in a 4-dimensional Hilbert space) whose
density matrix fulfils Tr$\left( \rho ^{2}\right) $ $<1/2$ may be actually
prepared. It is not difficult to show that such states never violate a Bell
inequality. That assumption is too strong and probably false, but it might
be the case that, even if the pure singlet state can be prepared, it evolves
in such a way that the spin correlation decreases with time, e.g. due to the
fact that in Dirac's theory the spin and the orbital angular momentum are
not separately conserved. (The situation is different with light, where the
polarization correlation is not lost with time \cite{BA},\cite{Weihs}). Or
it might be that the combination of spin correlation and position
correlation decreases in such a way that the Bell inequality can never be
violated. The reader is probably convinced that all these possibilities are
rather unlikely, but then I put him/her the challenge of explaining why it
is so extremely difficult to perform a loophole-free Bell experiment. I do
not have an explanation for the case of the spin-1/2 particles, but I have
one for the most common kind of experiments used to test Bell's
inequalities, namely those using correlated photon pairs produced in
parametric down conversion (PDC) \cite{CRS}. It is the case that most of the
''violations of a Bell inequality'' reported in the last 20 years have used
PDC (and suffer from the efficiency loophole).

In summary, I do not think that the question to be answered by future
experiments is whether quantum mechanics or local realism is true, but
whether there is a real contradiction between them or not.

\end{document}